\documentclass[english]{revtex4}
\usepackage[T1]{fontenc}
\usepackage[latin1]{inputenc}
\usepackage{graphicx}

\makeatletter

\usepackage{babel}
\makeatother
\begin{document}

\title{On the role of $C$ in $MgC_{x}Ni_{3}$, from the first-principles }

\author{P. Jiji Thomas Joseph and Prabhakar P. Singh}

\affiliation{Department of Physics, Indian Institute of Technology- Bombay, Mumbai
400076 India }

\begin{abstract}
The influence of vacancies in the $C$ sub-lattice of $MgCNi_{3}$,
on its structural, electronic and magnetic properties are studied
by means of the density-functional based Korringa-Kohn-Rostoker Green's
function method formulated in the atomic sphere approximation. Disorder
is taken into account by means of coherent-potential approximation.
Characterizations representing the change in the lattice properties
include the variation in the equilibrium lattice constants, bulk modulus
and pressure derivative of the bulk modulus, and that of electronic
structure include the changes in the, total, partial and $\mathbf{k}$-resolved
density of states. The incipient magnetic properties are studied by
means of fixed-spin moment method of alloy theory, together in conjunction
with the phenomenological Ginzburg-Landau equation for magnetic phase
transition. The first-principles calculations reveal that due to the
breaking of the $C$-$Ni$ bonds, some of the $Ni$ $3d$ states,
which were lowered in energy due to strong hybridization, are transfered
back to higher energies thereby increasing the itinerant character
in the material. The Bloch spectral densities evaluated at the high
symmetry points however reveal that the charge redistribution is not
uniform over the cubic Brillouin zone, as new states are seen to be
created at the $\Gamma$ point, while a shift in the states on the
energy scale are seen at other high symmetry points.
\end{abstract}
\maketitle

\section{Introduction}

Characterizations revealing the nature of pairing mechanism of the
$8$K perovskite superconductor $MgCNi_{3}$ \cite{Nature-411-54}
are at odds. The nuclear spin- lattice relaxation rate displays the
typical behaviour of isotropic $s$-wave superconductivity with a
coherence peak below the transition temperature $T_{C}$ \cite{PRL-87-257601}.
However, the specific heat \cite{Nature-411-54,PRB-70-174503,PhysicaC-388-563,PRB-67-052501,Supercond-17-274}
and resistivity \cite{Nature-411-54,PRB-64-132505,PRB-65-064534,PRB-67-052501,PRB-66-064510}
measurements imply a moderately coupled superconductor \cite{PRB-65-214518,PRB-67-052501,Supercond-17-274,J-SuperCond-15-485},
which is well supported by the tunneling experiments \cite{PRB-68-144510}
as well as theoretical calculations \cite{PRB-64-100508,PRB-64-180510}.
The penetration depth measurements however distinctly show a non $s$-wave
BCS feature at low temperature \cite{PRB-68-180502}. Hamiltonian
based model calculations, suggest $MgCNi_{3}$ to be a $d$- wave
superconductor \cite{SSC-122-269}. A two band model also have been
proposed \cite{PhysicaC-388-563,PRB-70-174503,condmat-0208367} to
reconcile the controversies in the experiments. Hall coefficient and
thermoelectric power data \cite{PRB-64-132505,PRB-67-052501} show
that the carriers are essentially electrons. However, it was also
suggested that the holes in $Ni$ $3d$ states could be responsible
for the transport properties, in analogy with the holes in the $O$
$2p$ states of perovskite oxide superconductors \cite{Nature-411-54}.
The constant scattering approximation also shows that the thermoelectric
power is hole-like above $10$K \cite{PRB-64-140507}. In fact the
multi- band model provides a consistent interpretation of the temperature
dependence of the normal resistivity, Hall constant and that of the
thermoelectric power \cite{condmat-0208367}.

The disparity in the experiments may arise due to the exactness in
terms of stoichiometry. Compositions namely $Mg_{1+z}C_{x}Ni_{3}$
with $0.75\leq z\leq1.55$ and $0.50\leq x\leq1.55$ are reported
\cite{PhysicaC-371-1}, when subjected to different synthesizing routes
(For a review, see \cite{JPCM-16-1237}). The variation then suggests
that the physical properties are intimately related to the temperature
which can be thought to control the materials composition and/or configuration
making it sensitive to the chemical nature of the compound as well
as the specific method used for crystal growth. For example, two different
phases- the $\alpha$ and $\beta$ have been synthesized by changing
the sintering temperature \cite{PhysicaC-371-1} with $\beta$ phase
being superconducting while the $\alpha$ phase remain non-superconducting.
The major difference between these two phases is in the exact $C$
content in the material. For high $C$ content, the scanning experiments
reveal the excess in terms of granules \cite{PRB-65-214518}. The
change from grain boundary to core pinning by these intra-granular
nano-particles near the superconducting transition temperature $T_{C}$
suggests that the arrangement of pinning sites in $MgCNi_{3}$ is
quite unique \cite{PRB-65-214518}. However, the physical properties
unveiled by the system is essentially a bulk property, not pertaining
to any interfaces or micro-structures. Nevertheless, the $T_{C}$
is sensitive to the $C$ content which decreases with decreasing $x$,
in $MgC_{x}Ni_{3}$ and disappears for materials where $x\leq0.9$
\cite{SSC-121-73}. The changes in the normal metal state properties
of $MgC_{x}Ni_{3}$ with respect to decreasing $x$ are what emphasized
in the present work.

The $C$ in the octahedral interstitial site of $MgC_{x}Ni_{3}$ has
two major roles to play. The spatially extended $C$ $2p$ orbitals
strongly hybridize with the $Ni$ $3d$, thus de-localizing the electronic
states. Delocalization leads to an overall reduction in its itinerant
character, rendering a definite non-magnetic state. Further, the presence
of $C$ expands the unit cell dimensions which favors soft $Ni$-
derived phonon modes \cite{PRB-64-100508}. A recent $C$- isotope
study \cite{cond-mat-0410504} and $B$ substitutions at $C$ sites
\cite{condmat-0412551} show that in addition to the dominant $Ni$
modes \cite{PRB-68-220504,PRB-69-092511}, certain $C$ modes are
also important in the materials pairing mechanism .

According to the microscopic theory of superconductivity, the occurrence
of vacancies may influence $T_{c}$ either through a modification
in the value of the density of states at Fermi energy $N(E_{F})$
or through modifying the electron-phonon coupling parameter, or both.
The electron-phonon coupling parameter is proportional to $M\left\langle \omega^{2}\right\rangle ,$
where $M$ is the mass of the transition metal and $\left\langle \omega^{2}\right\rangle $
is the average squared moment of the phonon spectrum. One expects
that vacancies in the $C$ sub-lattice of $MgC_{x}Ni_{3}$ can have
adverse effects on its lattice, electronic and other related properties.
it is shown in a previous report that the $N(E_{F})$ decreases, rather
abruptly, as an increasing function of vacancies \cite{PRB-68-024523}.
The decrease may be attributed to the disorder smearing of the electronic
bands, in particular to those which constitute the van-Hove singularity-like
feature just below the Fermi energy. However, it is possible that,
if not at at the Fermi energy, at least the upper valence band would
get enriched in states due to breaking of the $C$-$Ni$ bonds. This
may happen because some of the $Ni$ $3d$ states, which were lowered
in energy due to strong hybridization, would be transfered back to
the higher energies in proportion to the exact concentration of vacancies
in the $C$ sub-lattice. The redistribution can adversely affect the
electronic structure related properties such as magnetism. Earlier,
it was conjectured that since the hypothetical $MgNi_{3}$ shows a
definite ferromagnetic ground state, the disappearance of superconductivity
for $x\leq0.9$ alloys would be intimately related to the pair-breaking
effects \cite{JStrChem-43-168}. This, however, is inconsistent with
the experiments which detect no long-range magnetic ordering in $MgC_{x}Ni_{3}$
alloys \cite{SSC-121-73,PRB-68-024523}.

First- principles density functional based calculations are carried
out to study the changes in the equation of state parameters, viz
the equilibrium lattice constant, bulk modulus and its pressure derivative
as a function of $x$ in $MgC_{x}Ni_{3}$. The changes in the electronic
structure of the disordered $MgC_{x}Ni_{3}$ as a function of $x$
are described by means of total and sub-lattice resolved partial density
of states, calculated at their respective equilibrium lattice constants.
On the other hand the propensity of magnetism in $MgC_{x}Ni_{3}$
with respect to $x$ is studied by means of the fixed spin moment
method. The fit of the magnetic energy with magnetization to the Ginzburg-
Landau functional, and the variation of the Ginzburg- Landau coefficients
as a function of $x$ suggests that $C$ deficiencies can enhance
incipient magnetic properties. This is consistent with the self consistent
calculations which find a significant charge redistribution owing
to the transfer of certain $Ni$ $3d$ states, from the intermediate
energies to that close to the Fermi energy.

\section{Computational details}

The ground state properties are calculated using the Korringa-Kohn-Rostoker
(KKR) method \cite{ProgMater-27-1} formulated in the atomic-sphere
approximation (ASA) (Ref.\cite{Turek} and references therein) with
chemical disorder accounted by means of coherent-potential approximation
(CPA) \cite{PR-156-809}. For better refinements in the alloy energetics,
the ASA is corrected by the use of both the muffin-tin correction
for the Madelung energy \cite{PRL-55-600} and the multi-pole moment
correction to the Madelung potential and energy \cite{PRB-66-024201,PRB-66-024202}.
These corrections bring significant improvement in the accuracy by
taking into consideration the non-spherical part of the polarization
effects \cite{CMC-15-119}. The partial waves in the KKR-ASA calculations
are expanded up to $l_{max}=3$ inside the atomic spheres. The multi-pole
moments of the electron density have been determined up to $l_{max}^{M}=6,$
and then used for the multi-pole moment correction to the Madelung
energy. The exchange-correlation effects are taken into consideration
via the local density approximation (LDA) with Perdew and Wang parametrization
\cite{PRB-45-13244}. The core states have been recalculated after
each iteration. The calculations are partially scalar relativistic
in the sense that although the wave functions are non-relativistic,
first order perturbation corrections to the energy eigenvalues due
to the Darwin and the mass-velocity terms are included. Further, screening
constants $\alpha$ and $\beta$ were incorporated in the calculations,
following the prescription of Ruban and Skriver \cite{PRB-66-024201,PRB-66-024202}.
These values were estimated from the order($N$)- locally self-consistent
Green's function method \cite{PRB-56-9319} and were determined to
be $0.83$ and $1.18$, respectively. The atomic sphere radii of $Mg$,
$C$ and $Ni$ were kept as $1.404$, $0.747$, and $0.849$ of the
Wigner-Seitz radius, respectively. The vacancies in the $C$ sub-lattice
are modeled with the help of empty spheres, and their radius is kept
equal to that of $C$ itself. The overlap volume of the atomic spheres
was less than $15$\%, which is legitimate within the accuracy of
the approximation \cite{LMTO-1984}. The number of $\mathbf{k}$-points
for determining the total energies were kept in excess- $1771$ $\mathbf{k}$-points
in the irreducible wedge of the Brillouin zone. The convergence in
charge density was achieved so that the root-mean square of moments
of the occupied partial density of states becomes smaller than $10^{-6}$.
Numerical calculations of magnetic energy $\Delta E(M)$ for $MgC_{x}Ni_{3}$
are carried out at their self-consistently determined equilibrium
lattice constants using the fixed spin moment method \cite{JPF-14-129}.
In the fixed-spin moment method the total energy is obtained for given
magnetization $M$, i.e., by fixing the numbers of electrons with
up and down spins. In this case, the Fermi energies in the up and
down spin bands are not equal to each other because the equilibrium
condition would not be satisfied for arbitrary $M$. At the equilibrium
$M$ two Fermi energies will coincide with each other. The total magnetic
energy becomes minimum or maximum at this value of $M$.

\section{Results and discussion}

\subsection{Equation of state}

The estimation of the equilibrium lattice constant is a critical check
to the various approximations involved in the method. The KKR-ASA-CPA
method estimates the lattice constant of $MgCNi_{3}$ to be $7.139$
$a.u.$, which is $\sim1$\% less than the experimental value \cite{Nature-411-54}.
However, the value is consistent with an earlier full-potential based
method \cite{PRB-64-140507}. This shows that with the muffin-tin
correction \cite{PRL-55-600} the energetics of the material is described
more accurately, almost at par with the full-potential counterparts.
In the neutral spheres approximation \cite{PRB-49-1642}, in which
this correction is ignored, the equilibrium lattice constant for $MgCNi_{3}$
was determined to be $6.983$ $a.u.$

\begin{figure}[h]
\includegraphics[%
  scale=0.48]{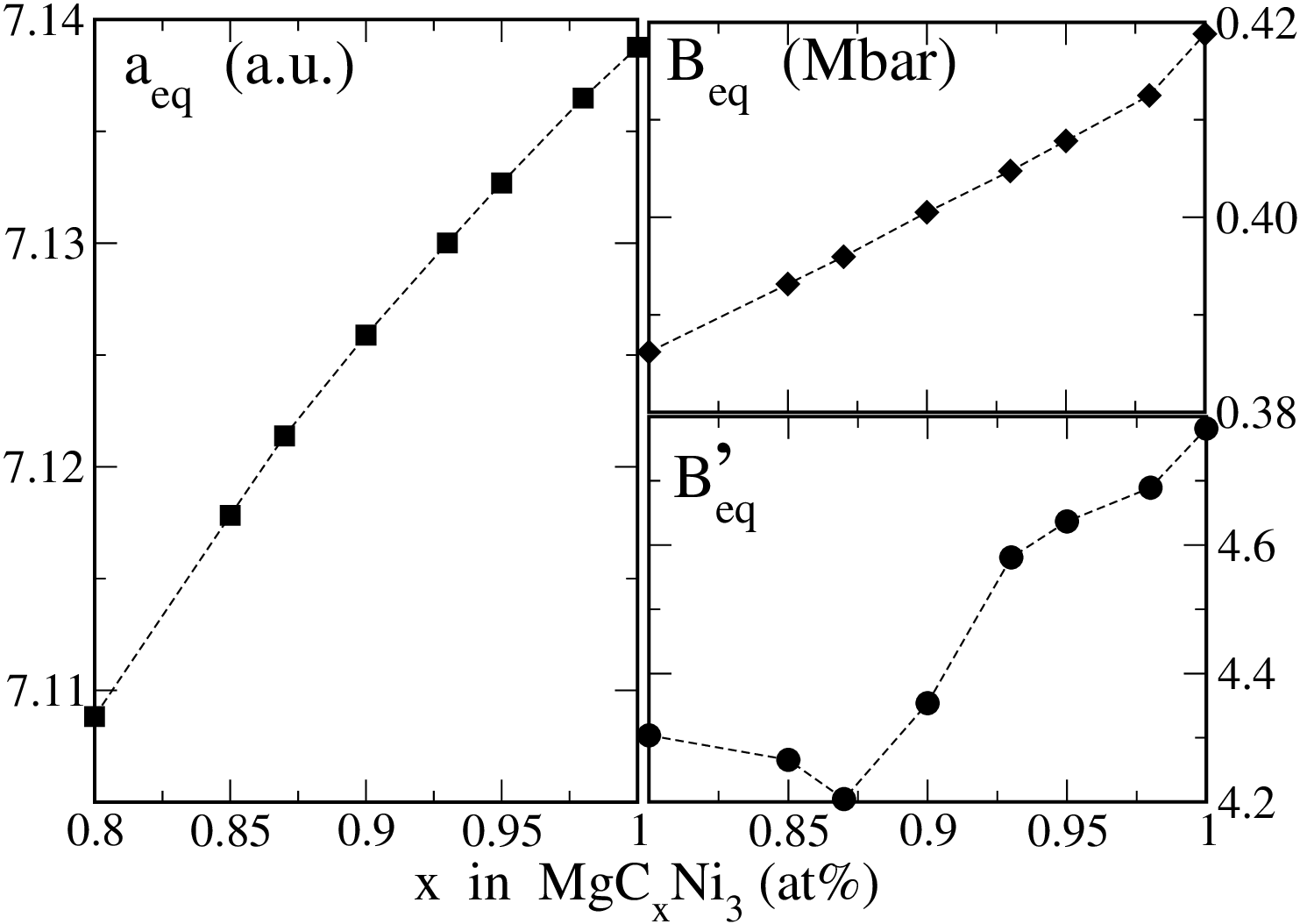}

\caption{\label{mgcxni3-eos}The change in the equation of state parameters
which include the equilibrium lattice constant ($a_{eq}$) in $a.u.$,
bulk modulus in Mbar and the pressure derivative of the bulk modulus,
as a function of $x$ in $MgC_{x}Ni_{3}$ alloys. }
\end{figure}

Total energy minimization en-route to the equilibrium lattice constants
$a_{eq}$, were extended to compositions studied in the range $0.8\leq x\leq1.0$
in $MgC_{x}Ni_{3}$. To estimate the bulk modulus $B_{eq}$ and its
pressure derivative $B_{eq}^{'}$, one may use the third order Birch-Murnaghan
equation of state \cite{JGR-57-227,FDES-140}. The variation of these
equation of state parameters as a function of $x$ are shown in Fig.\ref{mgcxni3-eos}. 

The structural parameters decrease as $x$ decreases in $MgC_{x}Ni_{3}$.
The rate of decrease in the equilibrium lattice constant with respect
to $x$ is found to be consistent with experiments \cite{SSC-121-73}.
The experiments find the rate of decrease as $0.187$ $a.u.$/$at$\%$C$
, while the present calculations find it to be $0.179$ $a.u.$/$at$\%$C$.
The decrease in $B_{eq}$ could be attributed to the fact that upon
creation of vacancies the material becomes some what hollow, making
it vulnerable to high compressibility. The change in $B_{eq}^{'}$
as a function of $x$ shows a minimum in the range of $0.85<x<0.90$.
The change in the slope of $B_{eq}^{'}$ empirically suggests a change
in the nature of chemical bonding as well as the lattice properties.
In the Debye approximation for isotropic materials, which assumes
a uniform dependence of the lattice frequencies with volume, one may
express the average phonon frequency in terms of $B_{eq}^{'}$ proportional
to $\,\frac{\delta\, ln\omega}{\delta\, lnV_{eq}}$, where $\omega$
is the average phonon frequency and $V_{eq}$ is the equilibrium volume
of the unit cell. Note that $V_{eq}$ decreases with decrease in $x$,
while $B_{eq}^{'}$ decreases and then increases. Such a behaviour
indicates that the properties associated with the lattice could be
different for $x<0.87$ and for $0.87<x<1.00$ alloys.

\subsection{Total and partial densities of states}

\begin{figure}[h]
\includegraphics[%
  scale=0.48]{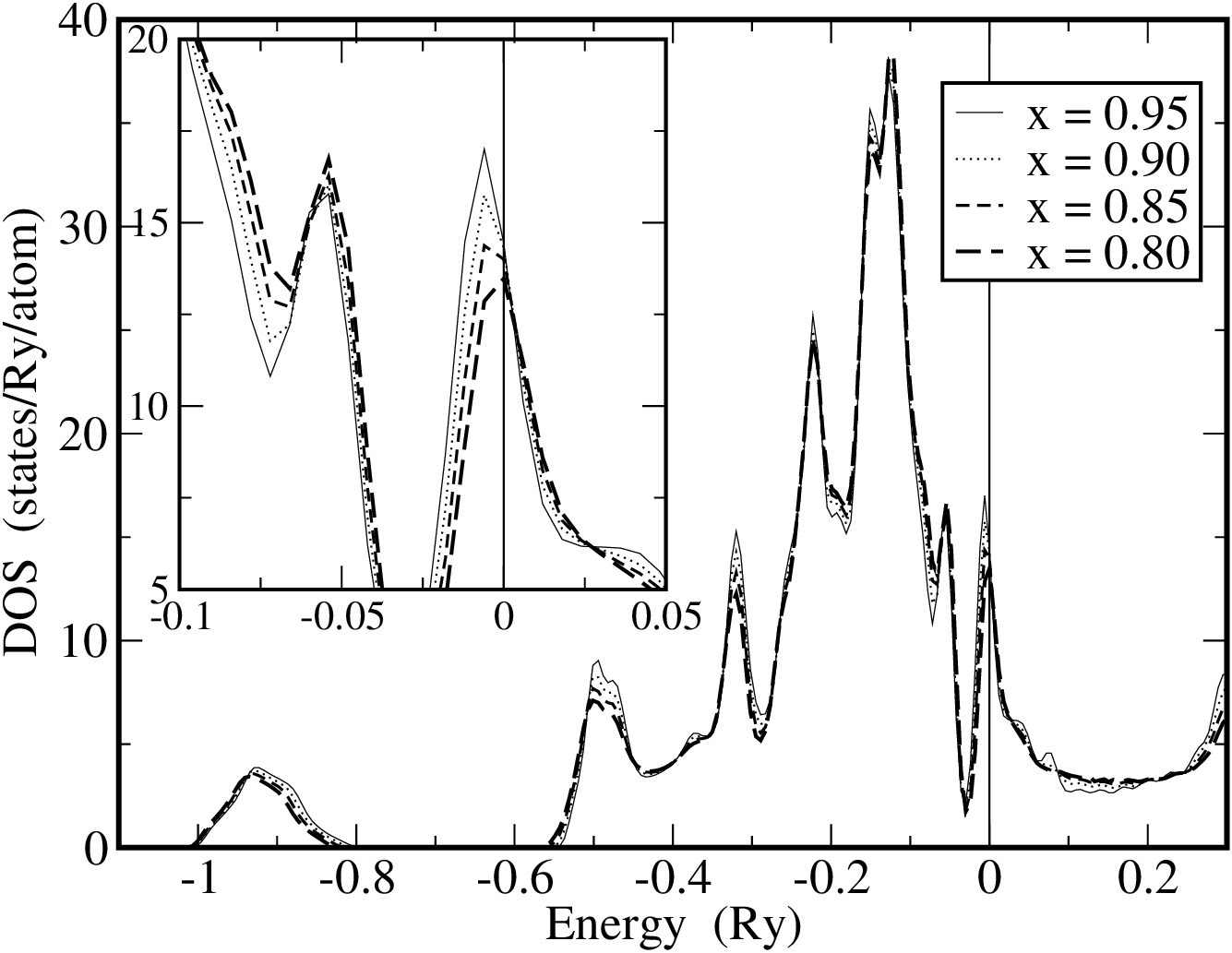}

\caption{\label{mgcxni3-dos}The change in the total density of state as a
function of $x$ in $MgC_{x}Ni_{3}$ alloys. The inset refers to a
blow up near the Fermi energy. }
\end{figure}

Upon creation of vacancies in the $C$ sub-lattice, some of the $C$
$2p$- $Ni$ $3d$ bonds break and and it can be expected that the
$3d$ states may redistribute in the higher energy spectrum constituting
the valence band. A possible way to examine this is to investigate
the alloy density of states as a function of $x$ in $MgC_{x}Ni_{3}$,
which is shown in Fig.\ref{mgcxni3-dos}. 

The present KKR-ASA-CPA calculations find $N(E_{F})$ for $MgCNi_{3}$
to be $14.56$ state/Ry-atom (or $5.35$ states/eV-cell). 
The present value
is in reasonable agreement with the previously reported values. 
For example, Szajek
reports the value as $5.26$ states/Ry-cell \cite{JPCM-L595}, Mazin
and Singh report as $4.99$ states/Ry-cell \cite{PRB-64-140507},
Shim $et$ $al$ as $5.34$ states/Ry-cell \cite{PRB-64-180510},
Rosner $et$ $al$ as $4.8$ states/Ry-cell \cite{PRL-88-027001}.
Dugdale and Jarlborg \cite{PRB-64-100508} report $N(E_{F})$ to be
$6.35$ and $3.49$ states/eV-cell for two different
band-structure methods with exchange-correlation effects considered
in the LDA and using the experimental lattice constant. These results
show that $N(E_{F})$ is indeed sensitive to the type of the electronic
structure method employed and also to the numerical values of the
parameters like that of the Wigner-Seitz radii and others. Note that
these differences are significant as they control the proximity to
magnetism in the Stoner model, as emphasized by Singh and Mazin \cite{PRB-64-140507}.

\begin{figure}[h]
\includegraphics[%
  scale=0.48]{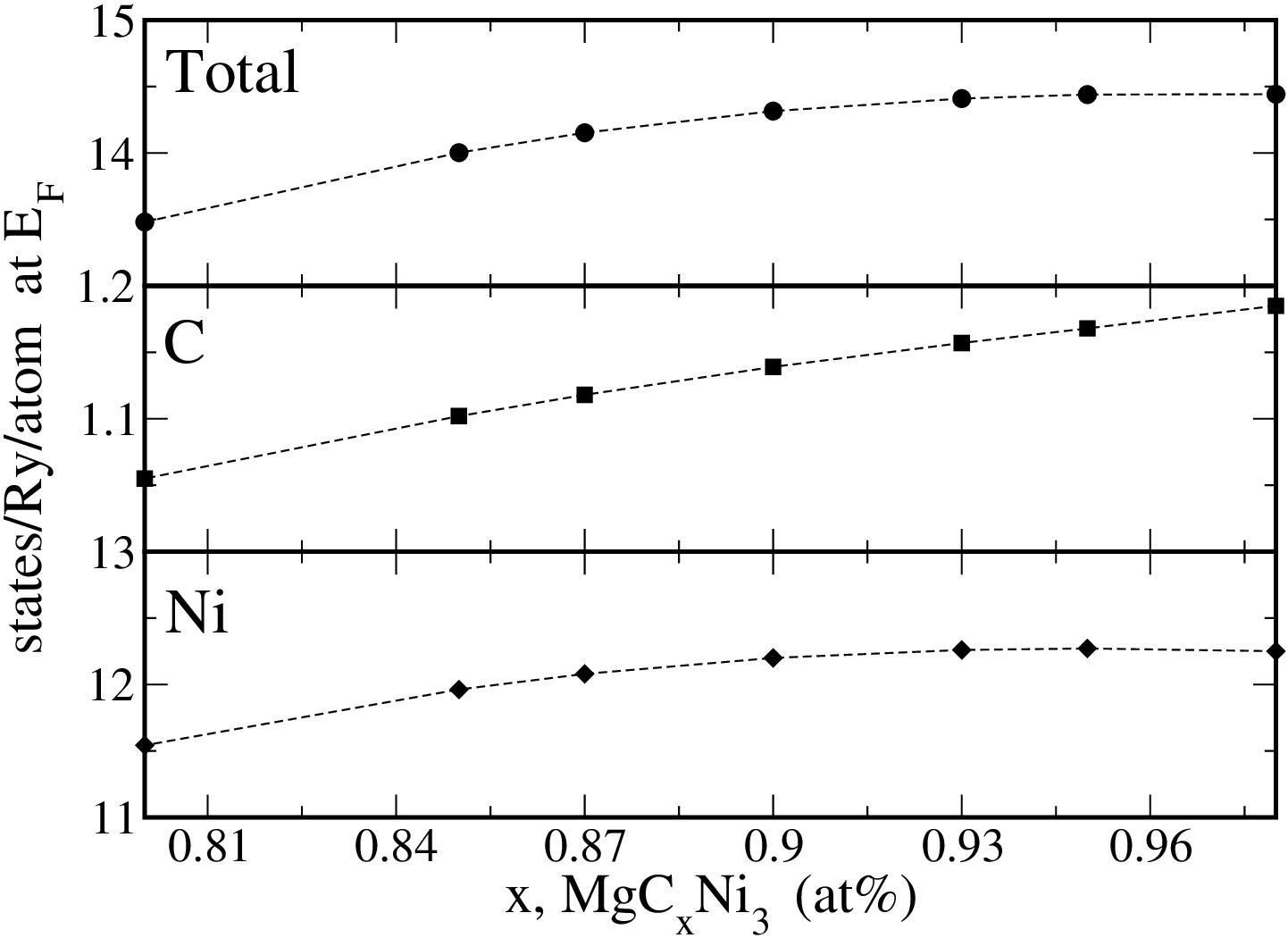}

\caption{\label{mgcxni3-nef}The change in the total, $C$ and $Ni$ contributions
to the density of states at the Fermi energy as a function of $x$
in $MgC_{x}Ni_{3}$ alloys. }
\end{figure}

As an decreasing function of $x$, a rigid band description to account
for the movement of $E_{F}$ fails for $MgC_{x}Ni_{3}$. The peak
characteristic of the $Ni$ $3d$ bands which lie just below the Fermi
energy, however, shows an insensitive change in its position on the
energy scale. Similar feature has also been reported by Rosner $et$
$al$ \cite{PRL-88-027001} for alkali metal substitutions at the
$Mg$ site in $MgCNi_{3}$. However, in $MgC_{x}Ni_{3}$, due to disorder
broadening of the bands, the $N(E_{F})$ is found to decreases. The
change in the total and sub-lattice resolved partial density of states
at the Fermi energy is shown in Fig.\ref{mgcxni3-nef}. The initial
change in $N(E_{F})$ as $x$ decreases from $1.0$ to $0.8$ is gradual,
however displays a rapid change when $C$ is reduced further. Such
a decrease in $N(E_{F})$ do not correspond to a rigid band picture.
Moreover, the $C$ $2p$ contribution to $E_{F}$ decreases monotonically
at the rate of $\sim0.709$ states/Ry-atom per $at$\% of $C$. while
$Ni$ $3d$ contribution remains more or less unaffected for the range
of alloys in $0.95<x<1.00$.

An earlier calculation based on CPA in the LMTO formalism finds a
drastic linear decrease in $N(E_{F})$ as a decreasing function of
$x$ \cite{PRB-68-024523}. The authors report $N(E_{F})$ to be $4.26$
states/eV-cell ($\simeq$$11.59$ states/Ry-atom) at $x$= $0.977$
which decreases quietly linearly to $3.14$ states/eV-cell ($\simeq$$8.54$
states/Ry-atom) at $x$= $0.85$. The present calculations, however,
observe a slightly different trend in the change of $N(E_{F})$ with
decreasing $x$ in $MgC_{x}Ni_{3}$. The $N(E_{F})$ at $x$=$0.98$
is calculated as $14.44$ states/Ry-atom and that at $x$=$0.85$
to be as $14.00$ states/Ry-atom. The discrepancy may be due to the
assumption of a particular lattice constant (experimental value) in
the previous TB-LMTO-CPA calculations \cite{PRB-68-024523}. 

As a prelude to understanding the overall shift in the bands as a
function of $x$ in $MgC_{x}Ni_{3}$, one may look at the shift in
the potential parameters, in particular the bottom of the band $^{B}B$
and the center of the band $^{B}C$. The center of the $Ni$ $3d$
band, $^{B}C_{Ni}$ shifts approximately by $+2.6$ mRy from $x=1.0$
to $0.8$ in $MgC_{x}Ni_{3}$. Note that here $+(-)$ represents the
movement of the corresponding potential parameter towards (away) the
Fermi energy. This is a clear revelation of the fact that the upper
valence band becomes enriched with states, as $x$ is decreased in
$MgC_{x}Ni_{3}$. Note that due to hybridization with the $C$ $2p$
states, some of the $Ni$ $3d$ states in $MgCNi_{3}$ are lowered
in energy. Upon creation of vacancies, a few of the $p-d$ bonds break,
and result in charge redistribution. We also observe that the $^{B}C_{Mg}$
shifts away from the Fermi energy by $-0.51$ Ry from $x=1.0$ to
$0.8$. In fact the $CNi_{6}$ octahedra is a covalently built complex
to which the strong electro-positive $Mg$ is thought to have donated
its outermost valence electrons. The crystal geometry suggests six
$Ni$ atoms as the first nearest neighbors to $C$ with a bond length
of $3.60$ $a.u.$, and eight $Mg$ atoms as its second nearest neighbors
with a bond distance of $6.25$ $a.u.$ in the unit cell. The $Mg$-$Ni$
bond length is $5.09$ $a.u.$, and as such for $Mg$'s the first
coordination shell comprises of twelve $Ni$ atoms. For $Ni$'s the
second nearest coordination shell carries four $Mg$ atoms. Hence,
the charge redistribution arising due to the breaking of the $p$-$d$
bonds are consistent with the fact that a larger fraction of the charge
is transfered back to the $Mg$ sub-lattice, in comparison with that
of the $Ni$ sub-lattice. 

The shift in the $^{B}C_{Ni}$ towards the Fermi energy indicates
an accumulation of $Ni$ $d$ states in the upper valence band which
then opens up two possibilities for the disappearance of superconductivity-
(i) The transfer of the states from low energy to higher energy states,
would increase the itinerant nature of electrons, thus enhancing the
possibility of spin-fluctuations, and (ii) if the $Ni$ $3d$ holes
are responsible for superconducting pairing mechanism, then these
states would be annihilated. To proceed further, one may require to
decompose the density of the states along various high symmetry points
and/ore directions of the cubic Brillouin zone.

\subsection{Bloch spectral density of states}

A convenient quantity to describe the $\mathbf{k}$-resolved density
of states of substitutionally disordered systems is the Bloch spectral
function $A_{B}(\mathbf{k},E)$, which is the number of states per
energy ($E$) and wavelength. In the case of pure metals the function
is simply a sum of delta functions either as function of $E$ at a
constant wave vector $\mathbf{k}$, or as a function of the wave-vector
$\mathbf{k}$ for constant value of the energy. For constant $E=E_{F}$,
where $E_{F}$ is the Fermi energy the positions of the peaks in $\mathbf{k}$-space
define the Fermi surface of the metal (see Refs.\cite{Turek,wienberger}
and references therein). The $\mathbf{k}$-space representation is
a good description of the electronic structure of the alloy although
strictly speaking there is periodicity only on the average. In alloys,
the Bloch spectral functions have their peaks lowered and broadened
due to disorder. Thus, a Fermi surface for the alloy is still defined
through the positions of the peaks but these peaks have a finite width. 

\begin{figure}[h]
\includegraphics[%
  bb=16bp 27bp 721bp 531bp,
  clip,
  scale=0.48]{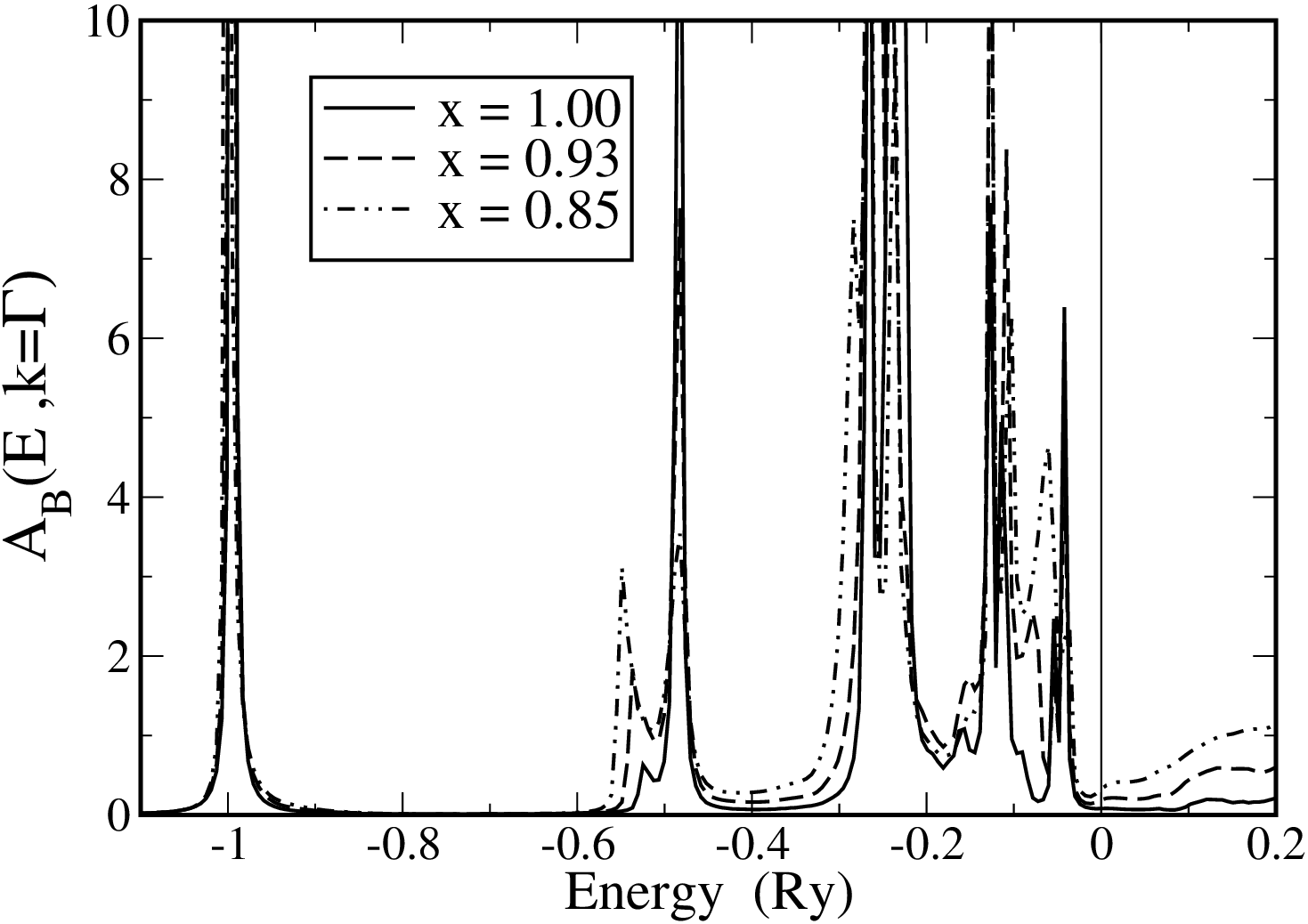}

\caption{\label{cx-bsf-G}The changes in the Bloch spectral density of states
$A_{B}(\mathbf{k},E)$ in arbitrary units, resolved at $\mathbf{k}=\Gamma$
for $MgC_{x}Ni_{3}$ alloys, using the LDA KKR-ASA-CPA method. The
vertical line through the energy zero represents the Fermi energy. }
\end{figure}

The states at the $\Gamma$ point of the Brillouin zone are shown
in Fig.\ref{cx-bsf-G}. The states corresponding to $-1$ Ry are primarily
composed of $Mg$ $3s$ states. The $C$ $2p$-$Ni$ $3d$ hybrids
form a complex in the energy range -$0.6<E<E_{F}$, where $E_{F}$
is taken as zero of the energy scale. As a function of vacancy concentration,
it is clear from Fig.\ref{cx-bsf-G} that a rigid-band picture is
violated. States are redistributed at this high symmetry point, with
a few of the $Ni$ $3d$ states lifted to higher energies, particularly
in the range $-0.1<E<E_{F}$. The peak corresponding to $-0.5$ Ry
has in it a component of $Mg$ $3s$ - $Ni$ $3d$ hybrid, which as
a function of $x$ lowers in energy. The sharp structure just below
the Fermi energy is characteristic of $Ni$ $3d$, which is less sensitive
to the vacancy concentration. What follows from Fig.\ref{cx-bsf-G}
is that the effects of vacancies are restricted to the energy range
$-0.6<E<E_{F}$, with a significant redistribution of states, leading
to an enhanced $Ni$ $3d$ character at $E_{F}$. There is a slight
increase in the density of states at $E_{F}$, which is far from any
rigid-band interpretation. 

\begin{figure}[h]
\includegraphics[%
  scale=0.48]{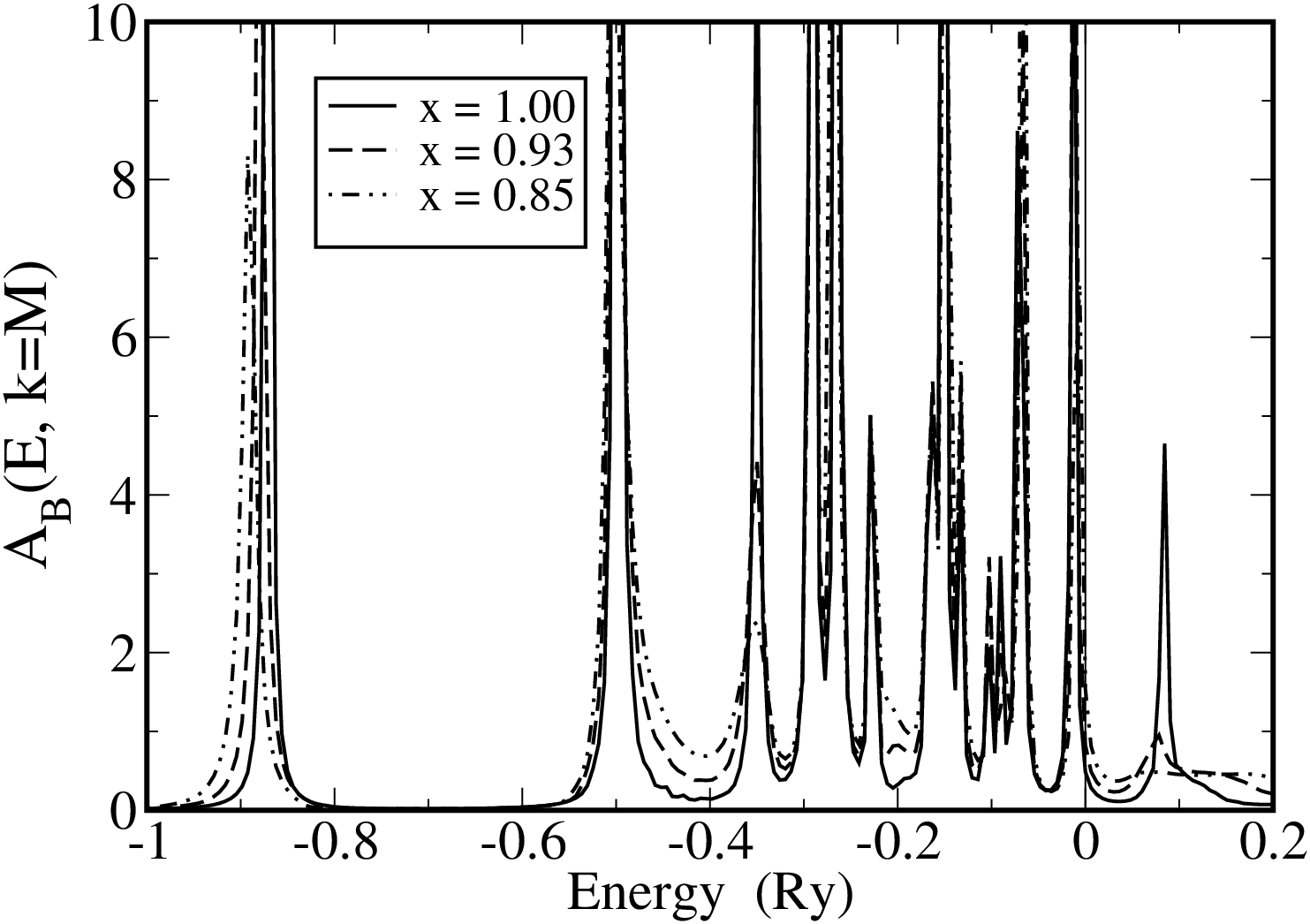}

\caption{\label{cx-bsf-M}The changes in the Bloch spectral density of states
$A_{B}(\mathbf{k},E)$ in arbitrary units, resolved at $\mathbf{k}=M$
for $MgC_{x}Ni_{3}$ alloys, using the LDA KKR-ASA- CPA method. The
vertical line through the energy zero represents the Fermi energy. }
\end{figure}

A flat band running close to the $M$ point of the Brillouin zone
has been mapped as a singularity in the DOS of $MgCNi_{3}$ \cite{PRL-88-027001}.
It has been emphasized that this singularity could induce magnetic
instability upon $0.5$ holes when introduced \cite{PRB-64-100508,PRL-88-027001}.
Fig.\ref{cx-bsf-M} shows the changes in the Bloch spectral function
for $MgC_{x}Ni_{3}$ alloys at the $M$ point of the Brillouin zone.
The states  near the Fermi energy are primarily $Ni$ $3d$ in character
and do not show any significant change with increase in vacancies.
The major effect of disorder is contained in two energy region, concentrated
at $-0.5$ and $-0.2$ Ry, respectively. These states result from
a strong hybridization effect of $Ni$ $3d$ and $C$ $2p$ bands.
Upon the creation of vacancies, the states concentrated around $-0.5$
Ry are raised in energy. The peak at $-0.4$ Ry lowers in height,
and the states are transfered to $-0.2$ Ry. The Fermi energy is almost
pinned at the foot of the sharp singularity, thus maintaining the
overall characteristic property of the alloy as a function of $x$
in $MgC_{x}Ni_{3}$ alloys. The peak that evolves towards the bottom
of the valence band is that of the $Mg$ $3s$ band, which moves slightly
towards lower energy. We note that in ASA formalism, the height of
the spectral peak at a particular energy do not exactly refer to the
states it holds, since the relevant matrix elements are ignored. However,
a comparison may still be useful, since the approximations involved
in the calculations are the same. 

\begin{figure}[h]
\includegraphics[%
  scale=0.48]{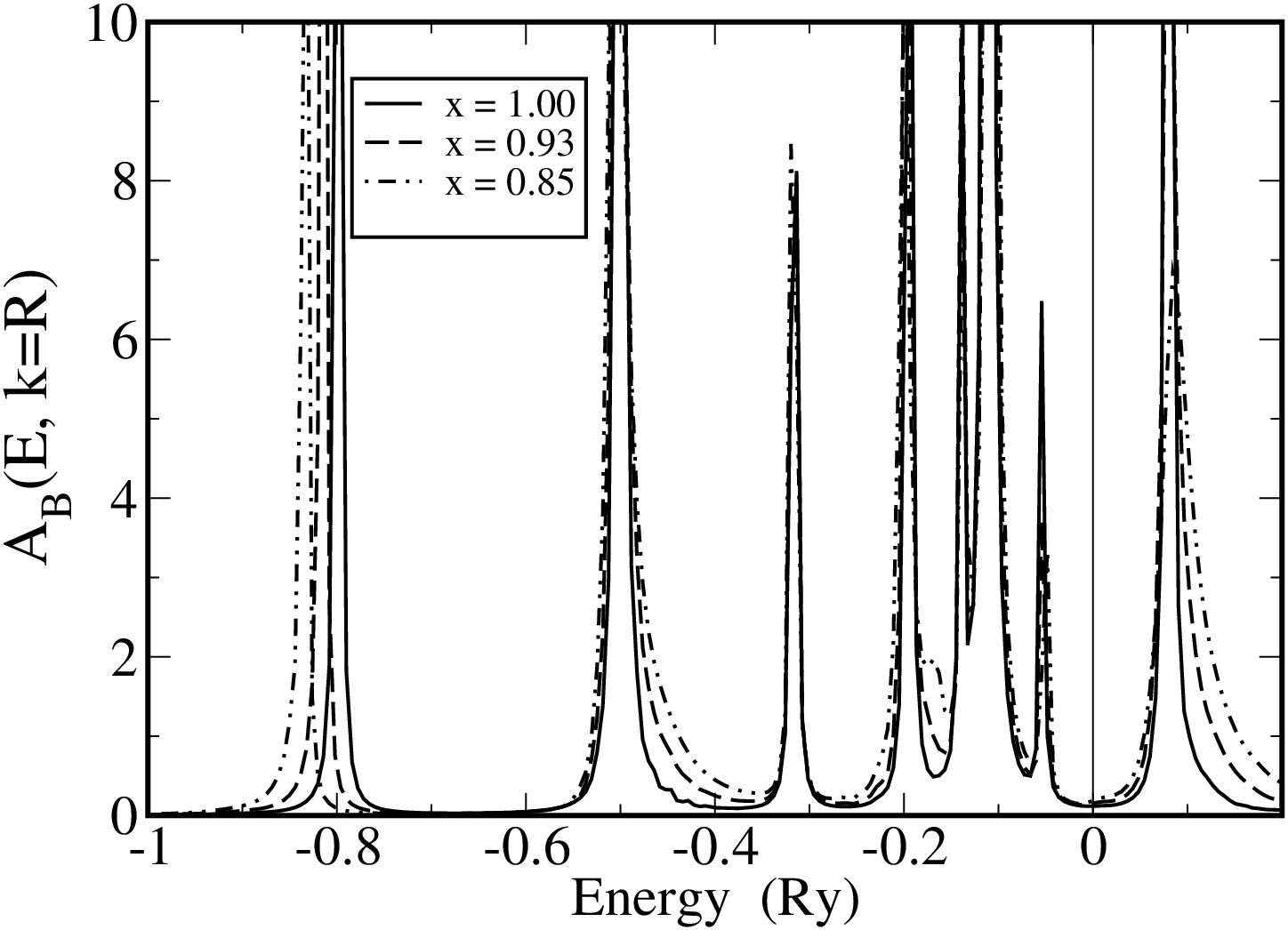}

\caption{\label{cx-bsf-R}The changes in the Bloch spectral density of states
$A_{B}(\mathbf{k},E)$ in arbitrary units, resolved at $\mathbf{k}=R$
for $MgC_{x}Ni_{3}$ alloys, using the LDA KKR-ASA- CPA method. The
vertical line through the energy zero represents the Fermi energy.}
\end{figure}

Figure \ref{cx-bsf-R} shows the redistribution of the states at the
$R$ point of the Brillouin zone in $MgC_{x}Ni_{3}$ as a function
of $x$. The states residing deep in energy are due to the $Mg$ $3s$
states, which move down as a function of increasing $x$. The major
effects take place where $C$ $2p$ bands are positioned, i.e., at
around $-0.45$ Ry and $-0.18$ Ry. The states in the energy range
$-0.18<E<E_{F}$ are strictly $Ni$ $3d$ in character which is least
affected. 

The importance of the $X$ point lies in the hole states it accommodates,
which is characterized by a peak just above the Fermi energy. As charge
redistributes, with the upper valence band becoming more $Ni$ $3d$
in character the hole states are however not annihilated. Instead,
the height of the peak decreases, showing a greater dispersion of
the bands. However, as for other high symmetry points, one can see
the distortion in those specific regions where $C$ $2p$ states dominate. 

In all cases, one finds that vacancies do not mimic a rigid band scenario.
Some of the low lying $Ni$ $3d$ states, due to hybridization with
$C$ $2p$ states, are restored back to the higher energies when the
bonds break. Charge transfer to the $Mg$ sub-lattice is seen to occur
via movement of the $Mg$ $3s$ peaks to lower energies as well as
an increase in the height of the peaks. As the charge redistributes,
the nature of the bonding could be adversely affected. The covalent
character of $MgC_{x}Ni_{3}$ then may decrease as an decreasing function
of $x$.

\subsection{Magnetic properties}

Incipient magnetism, due to delocalization of the $Ni$ $3d$ states
owing to the strong hybridization with the $C$ $2p$ are probable
in $MgCNi_{3}$. Following the Stoner criteria, it has been anticipated
that spin fluctuations may co-exist with superconductivity in $MgC_{x}Ni_{3}$
\cite{PRB-64-140507}. This however had been later confirmed in experiments
\cite{PRL-87-257601,PhysC-408-754}. 

\begin{figure}[h]
\includegraphics[%
  scale=0.48]{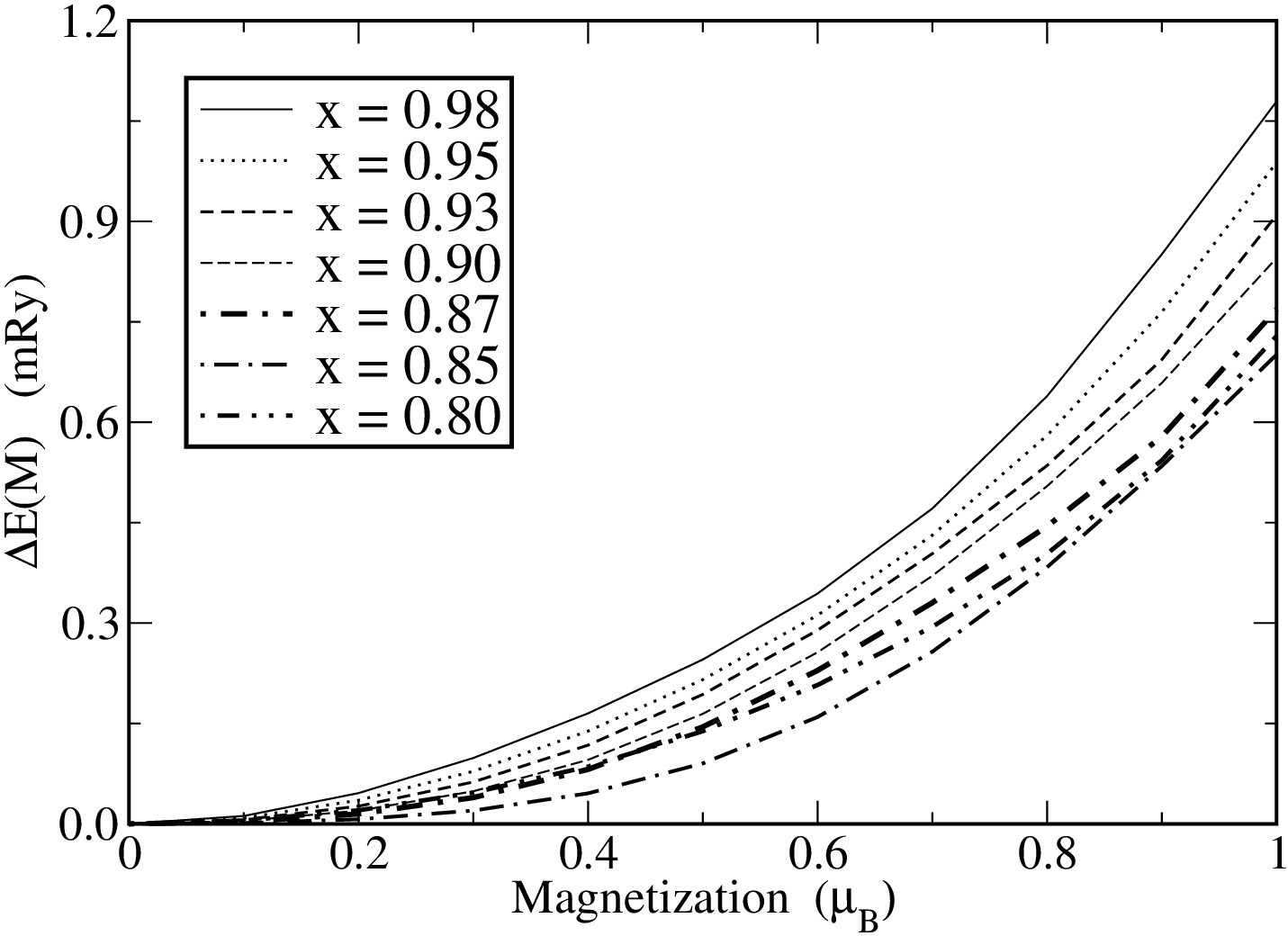}

\caption{\label{cx-fsmDeltaEM}The variation in the $\Delta E(M)$ in mRy
as a function of $M$ expressed in Bohr magnetons, of $MgC_{x}Ni_{3}$
alloys with $x$ as indicated in the leg box.}
\end{figure}

\begin{figure}[h]
\includegraphics[%
  scale=0.48]{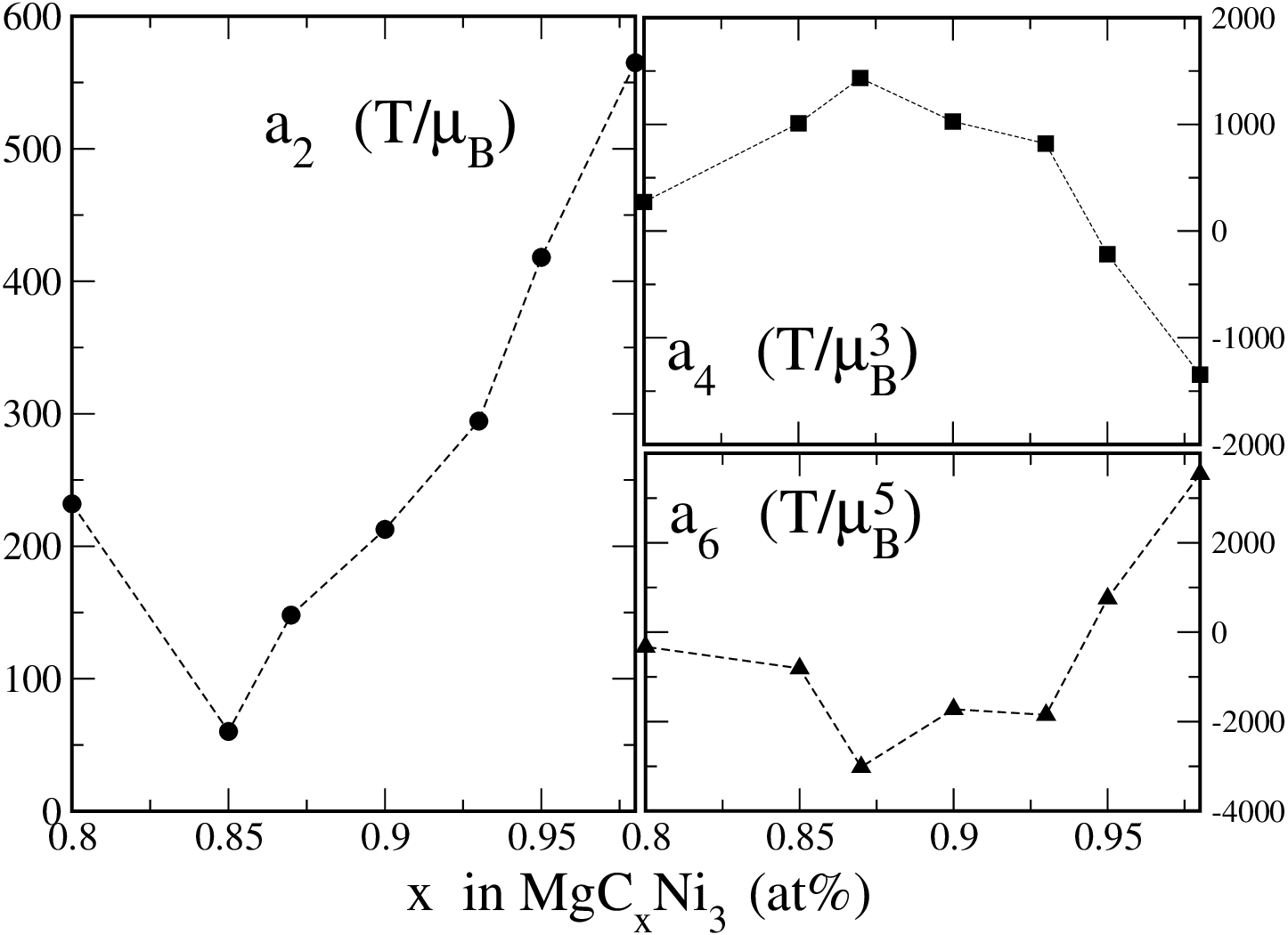}

\caption{\label{cx-coeffGL}The Ginzburg- Landau coefficients, $a_{2}$ expressed
in $(\frac{T}{\mu_{B}})$ and $a_{2}$ in $(\frac{T}{\mu_{B}^{3}})$
for $MgC_{x}Ni_{3}$ alloys as a function of $x$, from the LDA fixed
spin calculations. }
\end{figure}

Self consistent calculations, down to $x=0.8$ in $MgC_{x}Ni_{3}$,
do not find any magnetic solution. Both spin polarized and spin unpolarized
shows degeneracy in their total energies. To understand the change
in the incipient magnetic properties, as a function of $x$ in $MgC_{x}Ni_{3}$
alloys, one may then adopt the fixed-spin-moment method \cite{JPF-14-129}
by which the the energy difference between a magnetic state with magnetic
moment $M$ and the paramagnetic state, $\Delta E(M)$ (=$E(M)-E(0)$),
for given values of $M$ can be calculated. The calculated $\Delta E(M)$
is then fitted to the Ginzburg- Landau equation of form $\sum_{n>0}\,\,\frac{1}{2n}a_{2n}M^{2n}$.
The calculated results of $\Delta E(M)$ are shown in Fig.\ref{cx-fsmDeltaEM}.
The numbers shown in the figure denote the $C$ concentration. It
can be clearly seen that the curve of $\Delta E(M)$ is rather flat
near $M=0$ and the flatness increases as $x$ decreases. The curve
is fit to the form of a power series of $M^{2n}$ up to the term for
$n=3$, for the polynomial as mentioned above. The variation of the
coefficients, $a_{2}$ in units of $\frac{T}{\mu_{B}}$, $a_{4}$
in $\frac{T}{\mu_{B}^{3}}$, and $a_{6}$ in $\frac{T}{\mu_{B}^{5}}$
as a function of $x$, calculated at the respective equilibrium lattice
constants are shown in Fig.\ref{cx-coeffGL}. The propensity of magnetism
can be inferred from the sign of the coefficient which is quadratic
in $M$, i.e., $a_{2}$. The coefficient $a_{2}$ is the measure of
the curvature and is positive definite when the total energy minimum
is at $M=0$. This refers to the paramagnetic case. When $a_{2}$
becomes negative, it infers that there exists a minimum in the $E-M$
curve other than $M=0$ which then points to a ferromagnetic phase.
The higher order coefficients $a_{4}$ and $a_{6}$ are also significant
and control the variation. In fact, $a_{4}$ determines whether for
a system there exists a metastable state or not.

Consider a case where $a_{2}$ is small and positive while $a_{4}$
is relatively large and negative. In such a case, the calculations
extended to large values of $M$ would tend to show a metastable state
away from the total energy minimum. Often, calculations for large
values of $M$, implying large applied external fields, can lead to
ambiguous results. Hence, it is more accurate to carry out calculations
for smaller values of $M$ and use the above mentioned polynomial
function up to the minimum order, where the curve fits with sufficient
accuracy. 

\begin{figure}[h]
\includegraphics[%
  scale=0.48]{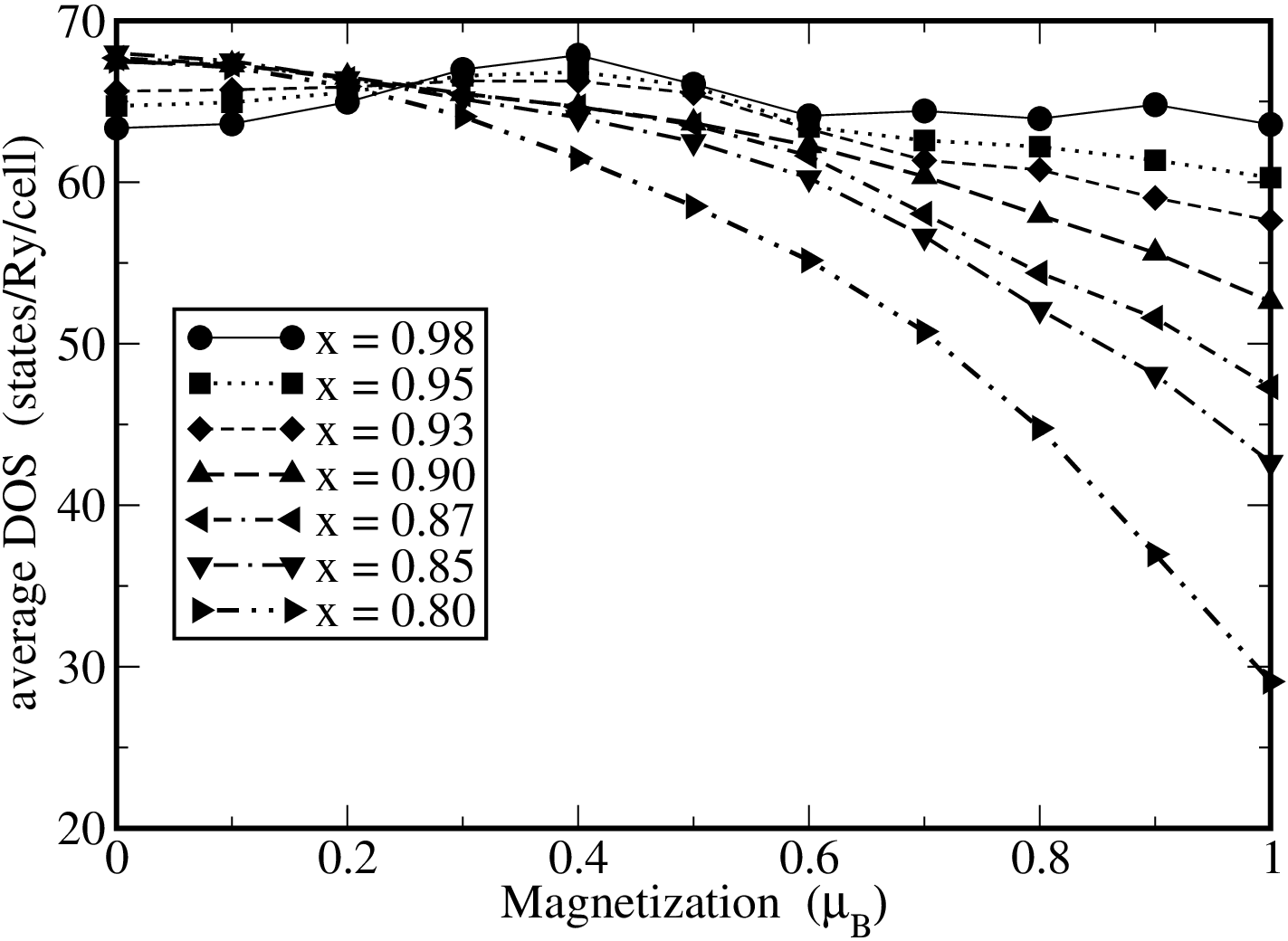}

\caption{\label{cx-avdos}The variation in the average DOS at $E_{F}$ as
a function of magnetization for different $x$ in $MgC_{x}Ni_{3}$
alloys from the LDA fixed spin calculations. }
\end{figure}

Fig. \ref{cx-coeffGL} shows that for $x\longrightarrow0.87$, from
the $C$ rich side, the alloys tend to enhance their latent magnetic
properties. One may also note that $a_{4}$ tends to become large
and negative. This however is complimented by an opposite variation
in the coefficient $a_{6}$. Hence, in the renormalized approach to
include corrections due to spin-fluctuations as like given in Ref.(\cite{PRB-59-9342}),
they would tend to cancel out, in proportion, preserving the trend
in the variation of $a_{2}$. Thus, it becomes more or less conclusive
within the approach that the incipient magnetic properties associated
with $C$ non-stoichiometry in $MgC_{x}Ni_{3}$ would increase as
a function of vacancies in the $C$ sub-lattice. If one considers
the proximity of $a_{2}\longrightarrow0$ as an indication of spin-fluctuations,
then the calculations within the approximation clearly show that effects
of spin-fluctuations would be dominant 
 in deciding the incipient magnetic properties of $MgC_{x}Ni_{3}$ 
as a function of decreasing
$C$. This is consistent with the movement of the center of the $Ni$
$3d$ bands towards Fermi energy, as well as a very little change
in the $N(E_{F})$ as a function of decreasing $x$, determined in
the self-consistent calculations. Since magnetism is controlled by
$N(E_{F})$, one may calculate the average density of states over
the two spin bands in the fixed-spin moment method. The variation
of this quantity as a function of $M$, for $x$ in $MgC_{x}Ni_{3}$
alloys is shown in Fig.\ref{cx-avdos}. Note that lattice relaxation
could be important \cite{PRB-67-064509} and particularly when $x$
increases in $MgC_{x}Ni_{3}$. However, the results that are discussed
are strictly for a case where one may find a rigid underlying lattice,
and also that the vacancies spread in the alloy is random. 

Thus, in general, it follows that as $x$ decreases in $MgC_{x}Ni_{3}$,
the propensity of magnetism increases. However, since $N(E_{F})$
decreases, the stoner criteria is less fulfilled, making the material
paramagnetic definite according to the itinerant models of magnetism.

\section{Summary and Conclusions}

The change in the equation of state parameters, density of states
and magnetic properties of $MgC_{x}Ni_{3}$ are studied by means of
the KKR-ASA-CPA method. Both lattice constant and bulk modulus decrease
with decreasing $x$, while the pressure derivative of the bulk modulus,
which is proportional to the average phonon frequency of the material
in the Debye approximation, is found to show an anomaly at about $x\sim0.87$.
The $N(E_{F})$ decreases, however not as expected by a rigid band
model. The Bloch spectral density evaluated at the $\Gamma$ point
shows creation of new states below $E_{F}$ while for other high symmetry
points they are shifted on the energy scale. The change in the coefficient
$a_{2}$ in the Ginzburg-Landau equation for magnetic phase transition
towards lower values suggests that incipient magnetic properties may
be enhanced as a decreasing function of $x$ in $MgC_{x}Ni_{3}$ alloys.
This is consistent with the fact that due to bond breaking, some of
the low lying $Ni$ $3d$ states are transfered to higher energies,
increasing the itinerancy of the material.

\section{Acknowledgements}

The authors would like to thank Hans L. Skriver and Andrei V. Ruban
for their KKR-ASA- CPA code used in the present work. One of us (PJTJ)
would like to thank Andrei V. Ruban for discussion on the theory and
implementation of code. Discussions with Dr. Igor I. Mazin on the
various aspects of materials properties are gratefully acknowledged.

\addcontentsline{toc}{chapter}{\numberline{}{Bibliography}}

\newpage

~

~

~

~

\begin{figure}[h]
\includegraphics[%
  scale=0.48]{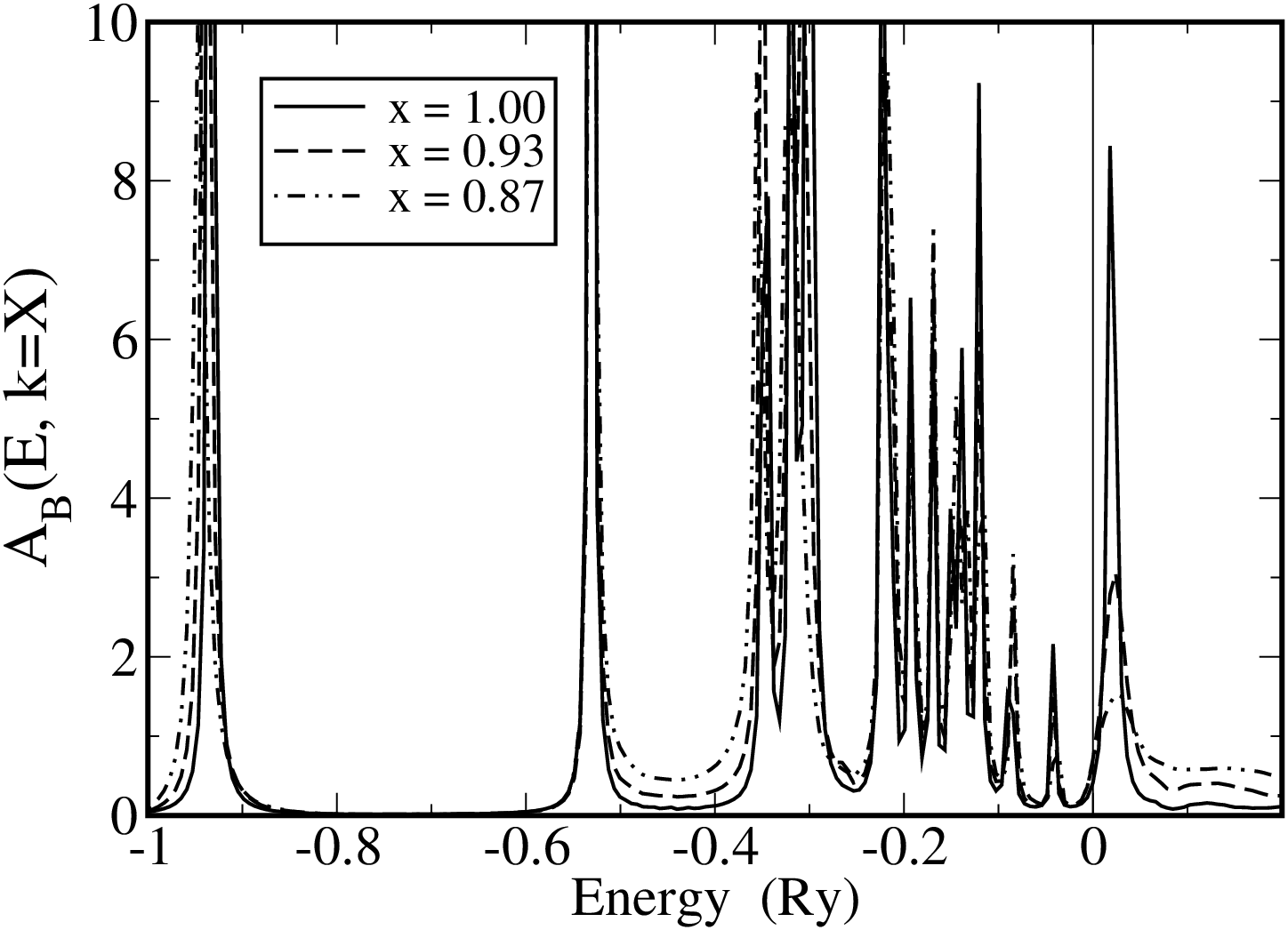}

\caption{\label{cx-bsf-X}The changes in the Bloch spectral density of states
$A_{B}(\mathbf{k},E)$ in arbitrary units, resolved at $\mathbf{k}=X$
for $MgC_{x}Ni_{3}$ alloys, using the LDA KKR-ASA- CPA method. The
vertical line through the energy zero represents the Fermi energy. }
\end{figure}

\end{document}